\documentclass[aps,pre,twocolumn,floatfix,nofootinbib,showpacs]{revtex4}
\usepackage{latexsym}
\usepackage{amsmath, amssymb}
\usepackage{epsfig}
\begin{document}
\newcommand{\be}{\begin{equation}}
\newcommand{\ee}{\end{equation}}
\title{Network analysis of online bidding activity}
\author{I. Yang$^1$, E. Oh$^1$ and B. Kahng$^{1,2}$}
\affiliation{{$^1$School of Physics and Center for Theoretical
Physics, Seoul National
University, Seoul 151-747, Korea}\\
{$^2$ Center for Nonlinear Studies, Los Alamos National
Laboratory, Los Alamos, New Mexico 87545.} }
\date{\today}
\begin{abstract}
With the advent of digital media, people are increasingly
resorting to online channels for commercial transactions. Online
auction is a prototypical example. In such online transactions,
the pattern of bidding activity is more complex than traditional
offline transactions; this is because the number of bidders
participating in a given transaction is not bounded and the
bidders can also easily respond to the bidding instantaneously. By
using the recently developed network theory, we study the
interaction patterns between bidders (items) who (that) are
connected when they bid for the same item (if the item is bid by
the same bidder). The resulting network is analyzed by using the
hierarchical clustering algorithm, which is used for clustering
analysis for expression data from DNA microarrays. A dendrogram is
constructed for the item subcategories; this dendrogram is
compared with a traditional classification scheme. The implication
of the difference between the two is discussed.
\end{abstract}
 \maketitle
\section {Introduction}
Electronic commerce (e-commerce) refers to any type of business of
commercial transaction that involves information transfer across
the Internet. Online auction, a synergetic combination of the
Internet supported by instantaneous interactions and traditional
auction mechanisms, has rapidly expanded over the last decade.
Owing to this rapid expansion and the importance of online
auctions, very recently researchers have begun to pay attention to
the various aspects of online
auctions~\cite{heck,rodgers,roth,heck2,yang1,yang2,born}.
According to recent studies based on empirical data obtained from
eBay.com, it was discovered that the online auction system is
driven by a self-organized process, involving almost all the
agents that participate in a given auction activity. For example,
the total number of bids placed for a single item or category and
the bid frequency submitted by each agent follow power-law
distributions~\cite{yang1}. Further, the bidding process occurring
in online auctions has been successfully described through the
stochastic rate equation~\cite{yang2}. Thus, understanding of the
bidding activities in online auctions is a highly attractive topic
for the statistical physics community.

The remarkable connection between {\em beer and diapers}
discovered in 1992 by Blischok et al.~\cite{beers} has
significantly improved profits. They analyzed the correlation
between items sold at a drug store during a particular time
interval between 5 p.m. and 7 p.m.. They found a strong
correlation between the two items, which had never been noticed by
the retailer earlier. This correlation arises from the fact that
fathers in families tend to buy beer when they are told by their
wives to buy diapers while returning home. This discovery, which
is considered as a pioneering work of data mining, compelled drug
stores to redesign their displays; this resulted in an increase in
beer sales.

In online auctions, most of the limitations hampering traditional
offline auctions, such as spatial and temporal constraint have
virtually disappeared. Thus, it would be interesting to
investigate how the bidding pattern of online auctions has changed
from the traditional one. On the other hand, recently,
considerable attention have been focused on complex network
problems as an interdisciplinary subject~\cite{rmp,mendes,siam}.
Diverse computational methods to find clusters within large-scale
networks have been introduced (for example, see
Refs.~\cite{girvan,newman,guimera,arenas,boguna}). Thus, by
combining these two issues, in this study, we investigate the
pattern emerging from the interactions between individual bidders
or items in online auctions by using the recently developed
network theory. The resulting network provides information on the
bidding pattern of individual bidders as well as the correlation
between different item subcategories. Moreover, we construct a
dendrogram for these subcategories and compare it with a
traditional classification scheme based on off-line transactions.
For the purpose, we use an algorithm applied for clustering
analysis for the expression data from a DNA microarray experiment
in biological systems~\cite{eisen}. The dendrogram thus obtained
is consumer-oriented, reflecting the pattern of an individual
bidder's activities. Thus, it can be used for increasing profits
by providing consumers with a link between the items, which should
interest the consumers.

Our study is based on empirical data \cite{yang1} collected from
http://www.eBay.com. The dataset comprises all the auctions that
ended in a single day, July 5, 2001, and includes 264,073
auctioned items grouped into 18 categories and 192 subcategories.
The number of distinct agents that participated in these
merchandize was 384,058.

\section {Topologies of bidder and item networks}

\begin{figure}
\centerline{\epsfxsize=6cm \epsfbox{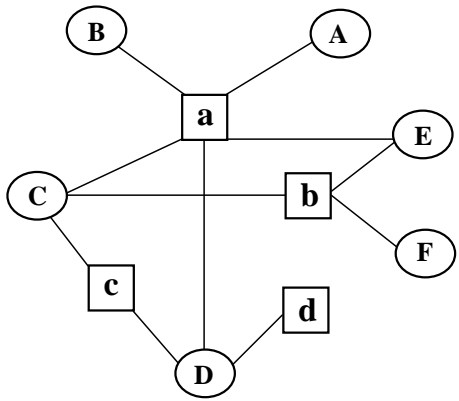}} \caption{A
schematic illustration of a bipartite network of an online
auction. Bidders and items are represented by ellipses with
\{A,B,\dots F\} and squares with \{a,b,c,d\}, respectively.
Bidders A and B are connected via item a which they bid for. Items
a and b are connected via bidders C and E who bid for both items a
and b.} \label{FIG:AUCTION_NET}
\end{figure}

\begin{figure}
\centerline{\epsfxsize 6cm \epsfbox{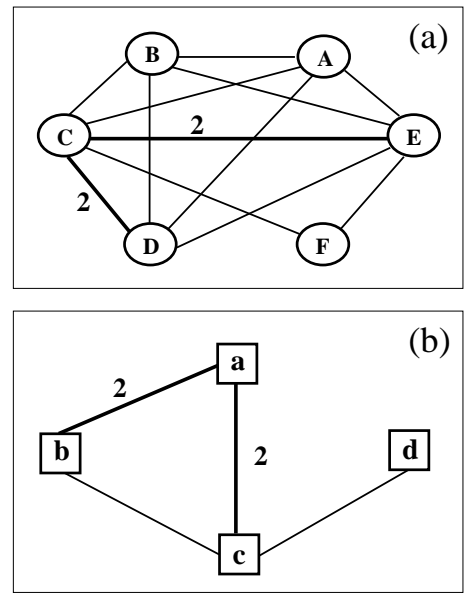}} \caption{Bidder
network (a) and item network (b) converted from the bipartite
network shown in Fig.~1. Thick edges have weight $2$, while the
other edges have unit weight for both (a) and (b).}
\label{FIG:AUCTION_NET2}
\end{figure}

The data contain the information on which bidder bids for which
item via their unique user ID. Thus, we can construct a bipartite
network comprising two disjoint sets of vertices, bidders and
items, as shown in Fig.~\ref{FIG:AUCTION_NET}. The bipartite
network can be converted to a single species of network such as
the bidder or the item network, as shown in
Figs.~\ref{FIG:AUCTION_NET2}(a) and \ref{FIG:AUCTION_NET2}(b),
respectively. The bidder and item networks can have edges with
weight. For example, bidders C and D in Fig.~\ref{FIG:AUCTION_NET}
are connected twice through items a and c. Hence, the edge between
C and D has weight 2. Similarly, items a and b are connected twice
through bidders C and E. Thus, the edge between vertices a and b
in the item network has weight 2. Statistics describing the
topology of the entire network and the giant component of the
bidder and the item network are listed in Table
\ref{TBL:TOTAL_NET} and Table \ref{TBL:GIANT_NET}, respectively.

\begin{table}
\begin{tabular}{c|c c c}
\hline\hline
 & $N$ & $L$ & $C_{\rm iso}$ \\
\hline
 ~~Bidder network~~ &~~338,478~~&~~1,208,236~~&~~22,883 \\
\hline
 Item network &   122,827& 813,687 & 3,851\\
\hline
\end{tabular}
\caption{The numbers of vertices $N$, edges $L$, and isolated
clusters $C_{\rm iso}$ for the bidder and the item networks.}
\label{TBL:TOTAL_NET}
\end{table}

\begin{table}
\begin{tabular}{c|c c c c}
\hline\hline
 & $N$ & $L$ & $\langle k \rangle$  &  $\langle d\rangle$ \\
\hline
~~Bidder network~~ & ~~267,414~~ & ~~2,245,794~~ & ~~8.4~~ & ~~8.15~~ \\
  &      (79\%) & (93\%) &   &   \\
 \hline
Item network &    112,240 & 695,281 & 12.4 & 7.69 \\
  &       (91\%) & (85\%) &   &   \\
 \hline
\end{tabular}
\caption{Statistics of the giant component of the bidder and item
networks. The number of vertices is denoted by $N$; edges, $L$;
mean degree, $\langle k \rangle$; and mean distance between two
vertices, $\langle d \rangle$.} \label{TBL:GIANT_NET}
\end{table}

\begin{figure}
\epsfxsize 8cm \epsfbox{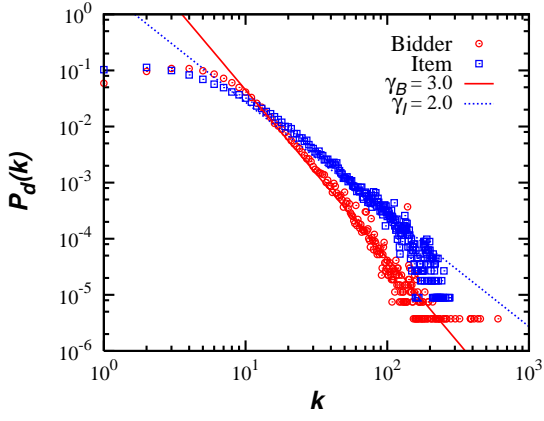} \caption{Degree distribution
$P_d(k)$ as a function of degree $k$. Both display power-law
behaviors $P_d(k) \sim k^{-\gamma}$ with $\gamma_B \approx 3.0$
for the bidder network and $\gamma_I \approx 2.0$ for the item
network. Solid lines are guidelines with slopes of 3.0 and 2.0 for
the bidder and item networks, respectively. }\label{FIG:PK}
\end{figure}

Next, we characterize the structure of the bidder and item
networks. First, we regard each network as a binary network,
neglecting the weight of each edge. The network configuration can
be described by the adjacent matrix $\{a_{ij}\}$; its component is
1 when two vertices $i$ and $j$ are connected and 0 otherwise.
Then, degree $k_i$ of vertex $i$ is $k_i=\sum_j^N a_{ij}$, which
is the number of edges connected to it. We find that the degree
distribution exhibits a power-law behavior asymptotically for both
the bidder and item networks, $P_d(k)\sim k^{-\gamma}$. The degree
exponent $\gamma$ is estimated to be $\gamma_B\approx 3.0$ for the
bidder network and $\gamma_I\approx 2.0$ for the item network, as
shown in Fig.~\ref{FIG:PK}.

\begin{figure}
\epsfxsize 7cm \epsfbox{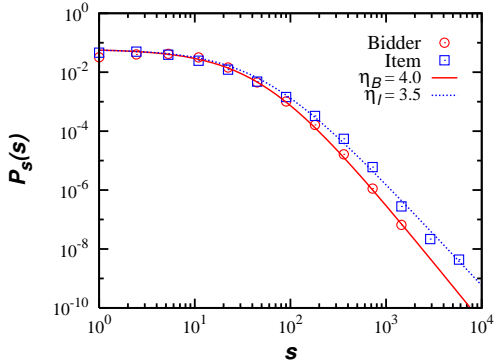} \caption{Strength distributions
$P_s(s)$ as a function of strength $s$ for the bidder and item
networks. Asymptotically, they display a generalized power-law
behavior $P_s(s)\sim (s+s_0)^{-\eta}$. The exponent is estimated
to be $\eta_B \approx 4.0$ for the bidder network and $\eta_I
\approx 3.5$ for the item network. $s_0=51$ is used for the bidder
network and $s_0=52$ for the item network.} \label{FIG:PS}
\end{figure}

\begin{figure}
\epsfxsize 7cm \epsfbox{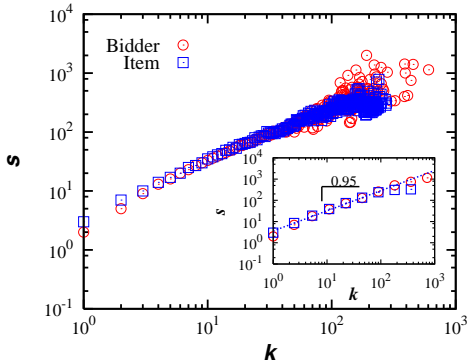} \caption{The relation between
strength $s$ and degree $k$ of each vertex. They show an almost
linear relationship, $s \sim k^{\zeta}$ with $\zeta \approx 0.95$
for both the bidder ($\circ$) and the item ($\square$) network.
Inset: Replot of $s$ vs. $k$ using the log-bin data.}
\label{FIG:S_K}
\end{figure}

Second, strength $s_i$ of vertex $i$ is the sum of the weights of
each edge connected to it. That is, $s_i=\sum_{j}^N a_{ij}w_{ij}$,
where $w_{ij}$ is the weight of the edge between vertices $i$ and
$j$. The strength distributions of the bidder and item networks
also exhibit power-law behaviors asymptotically as $P_s (s)\sim
(s+s_0)^{-\eta}$ where $\eta_B\approx 4.0$ for the bidder network
and $\eta_I \approx 3.5$ for the item network, as shown in
Fig.~\ref{FIG:PS}. $s_0$ is constant. Strength and degree of a
given vertex exhibit an almost linear relationship $s(k)\sim
k^{\zeta}$ with $\zeta\approx 0.95$; however, large fluctuations
are observed for large $k$ in Fig.~\ref{FIG:S_K}.

Third, we measure the mean nearest-neighbor degree function
$\langle k_{\rm nn}\rangle (k)$. The mean degree of the
nearest-neighbor vertices of a given vertex $i$ with degree $k$ is
measured as follows:
\begin{equation}
k_{i,\rm nn}=(\sum_j a_{ij}k_j)/k_i. \label{eq:kinn}
\end{equation}
The average of $k_{i,\rm nn}$ over the centered vertex with degree
$k$ is taken to obtain $\langle k_{\rm nn}\rangle (k)$. For the
weighted network, formula (\ref{eq:kinn}) is replaced following
the formula~\cite{vespignani}:
\begin{equation}
k_{i,\rm nn}^{(w)}=\frac{1}{s_i}\sum_j a_{ij} w_{ij} k_j.
\end{equation}
From this equation, $\langle k_{\rm nn}^{(w)}\rangle (k)$ can be
similarly obtained. It is found that the functions $\langle k_{\rm
nn}\rangle (k)$ and $\langle k_{\rm nn}^{(w)}\rangle (k)$ increase
with the degree $k$ of the centered vertex for both the bidder and
item networks irrespective of the binary or weighted versions.
That is, both the networks are assortatively mixed, implying that
{\em active bidders tend to simultaneously bid for common items,
thereby attractive items are also connected via such active
bidders.}

\begin{figure}
\epsfxsize 9cm \epsfbox{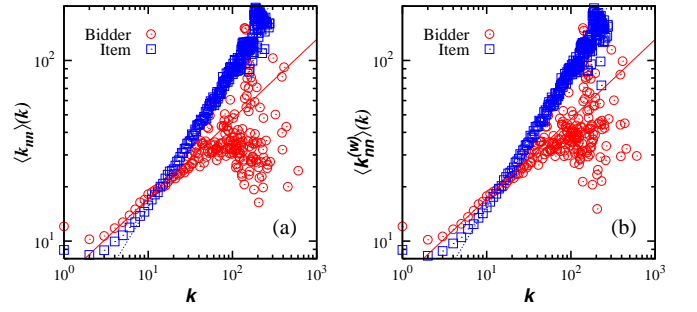} \caption{The mean
nearest-neighbor degree function $\langle k_{\rm nn}\rangle (k)$
($\circ$) and its weighted version $\langle
k_{nn}^{(w)}\rangle(k)$ ($\square$) as a function of the degree
$k$ of a centered vertex for the bidder network (a) and the item
network (b). Solid line, obtained from a least-square-fit, has a
slope of $0.44$ for the bidder network (a) and $0.77$ for the item
network (b). Both the networks are assortatively mixed.}
\label{FIG:K_KNN}
\end{figure}

\begin{figure}
\epsfxsize 9cm \epsfbox{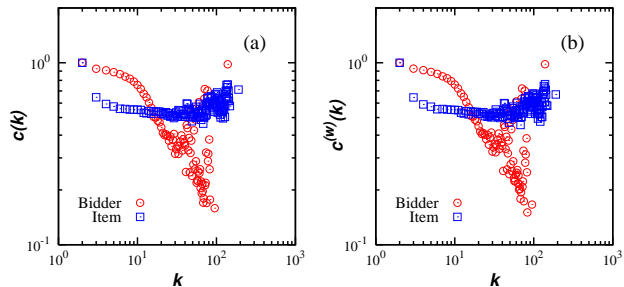} \caption{Average clustering
coefficient functions $c(k)$ ($\circ$) and $c^{(w)}(k)$
($\square$) as a function of degree $k$ for the bidder network (a)
and the item network (b). The result implies that the bidder
network is hierarchically organized, whereas the item network is
almost random.} \label{FIG:K_CK}
\end{figure}

Fourth, the local clustering coefficient $c_i$ is the density of
transitive relationships, and is defined as the number of
triangles formed by its neighbors, which are cornered at vertex
$i$, divided by the maximum possible number of neighbors,
$k_i(k_i-1)/2$. That is,
\begin{equation}
c_i = \frac{2}{k_i(k_i-1)} \sum_{j,h} a_{ij}a_{ih}a_{jh}.
\end{equation}
The average of $c_i$ over the vertices with degree $k$ is called
the clustering coefficient function $c(k)$. For weighted networks,
a similar clustering coefficient $c_i^{(w)}$ is
defined~\cite{vespignani} as
\begin{equation}
c_i^{(w)}=\frac{1}{s_i(k_i-1)}\sum_{j,h}\frac{w_{ij}+ w_{ih}}{2}
a_{ij}a_{ih}a_{jh}.
\end{equation}
The average of $c_i^{(w)}$ over the cornered vertices with degree
$k$ is similarly defined and denoted as $c^{(w)}(k)$. For the
bidder network, the clustering coefficient functions $c(k)$ and
$c^{(w)}(k)$ decrease with respect to $k$ as shown in
Fig.\ref{FIG:K_CK}(a); they exhibit large fluctuations for large
$k$, implying that the bidder network is hierarchically organized.
For the item network, however, both $c(k)$ and $c^{(w)}(k)$ are
almost independent of $k$, which is shown in
Fig.~\ref{FIG:K_CK}(b); this implies that the network is almost
randomly organized. Such behaviors are observed irrespective of
whether the networks are binary or weighted.

\section {Cluster identification}

By using network analysis, individual elements can be classified
into clusters. Here, we apply the hierarchical agglomeration (HA)
algorithm, which was introduced by Clauset {\it et
al.}~\cite{clauset}, to the item network containing 264,073 items.
In particular, the algorithm is useful for a system containing a
large number of elements. Clusters identified using this analysis
are compared with traditional subcategories established based on
off-line transactions. The obtained difference can be used for
reorganizing a dendrogram with regard to item subcategories; this
difference reflects the pattern of online bidding activities.

To realize this, we first store the topology of the item network
by using the adjacent matrix $\{a_{ij}\}$. By maintaining this
information, we delete all the edges, thereby leaving $N$ isolated
vertices. At each step, we select one edge from the stored
adjacent matrix, which maximizes a change in the modularity,
defined as
\begin{equation}
Q=\sum_{\alpha} e_{\alpha \alpha}-a_{\alpha}^2,
\end{equation}
where $e_{\alpha \alpha}$ is the fraction of the edges that
connect the vertices within cluster $\alpha$ on both the ends of
each edge, and $a_{\alpha}$ is the fraction of edges attached on
one end or both the ends to vertices within cluster $\alpha$. The
selected edge is eliminated from the stored matrix. We continue
this edge-adding process until the modularity becomes maximum. We
find that the modularity reaches the value $Q_{\rm max}\approx
0.79$ for the item network and $Q_{\rm max}\approx 0.83$ for the
bidder network; this implies that both the networks are extremely
well categorized. We recognize 1,904 and 870 distinct clusters in
the bidder and item networks, respectively. The cluster sizes, the
number of vertices of each module, are not uniform. The
cluster-size distributions for both networks, even though large
deviations exist for a large cluster size $M$, exhibit fat-tail
behaviors such that $P_m(M)\sim M^{-\tau}$ with $\tau_B \approx
2.2$ and $\tau_I \approx 2.1$ roughly. The exponents are estimated
from the data in the region with small $M$.

\begin{figure}
\epsfxsize 7cm \epsfbox{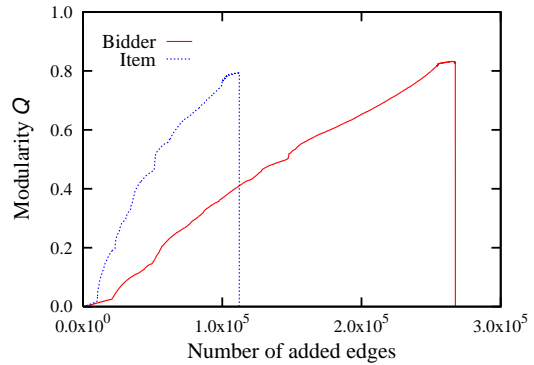} \caption{The evolution of
modularity $Q$ by using the edge-adding process. The $x$ axis
represents the number of edges added. The maximum value obtained
is estimated to be $Q_{max}=0.83$ for the bidder network (solid
line) and $Q_{max} = 0.79$ for the item network (dotted line).}
\label{FIG:MODULARITYQ}
\end{figure}

\begin{figure}
\epsfxsize 7cm \epsfbox{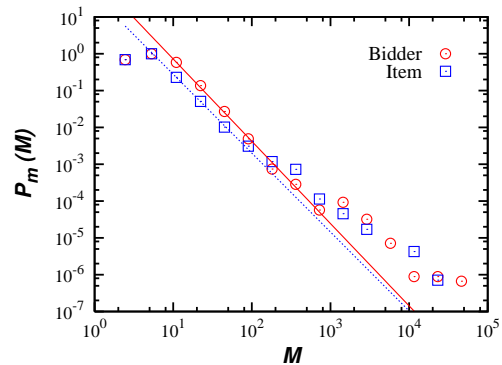} \caption{The cluster-size
distributions for the bidder and item networks, identified using
the HA algorithm. The distributions follow the power law,
$P_m(M)\sim M^{-\tau}$ with $\tau_B\approx 2.2$ and $\tau_I\approx
2.1$. The exponents are estimated from the region with the data in
small $M$. Solid and dashed lines are guidelines. The presented
data are log-binned. Raw data in the region with large $M$ are
sparse.} \label{FIG:DENSITYOFCOMMUNITY}
\end{figure}

\section{Dendrogram based on online transactions}

\subsection{Closeness}

In this section, we focus on the item network. We have identified
870 distinct clusters by using the clustering algorithm. Among
them, 49 clusters contain more than 100 items within each cluster.
On the other hand, according to the traditional classification
scheme, items in the eBay auction are categorized into 18
categories which contains 192 subcategories. Obviously the
clusters that we found are not equivalent to these categories or
subcategories. Thus, our goal is to construct a new dendrogram, a
hierarchical tree, among 192 subcategories based on the closeness
between the obtained clusters and the existing subcategories.

\begin{figure*}
\center \epsfxsize 15cm \epsfbox{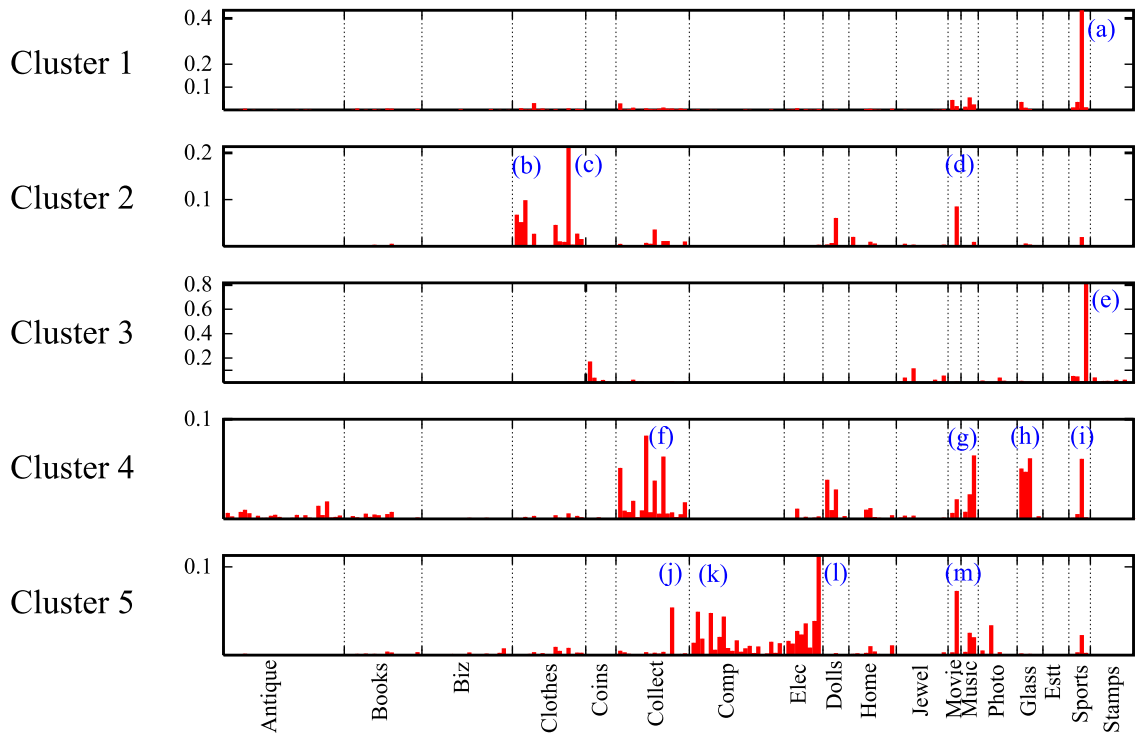} \caption{The closeness
between the clusters and subcategories. The $x$-axis represents
the 192 subcategories and the $y$-axis represents the closeness.
For the largest cluster (cluster 1), subcategory (a), Sports and
Goods, exhibits the largest closeness. This result indicates that
the main fraction of items in cluster 1 originates from
subcategory (a), even though small fractions of items exist from
other subcategories. For cluster 2, the subcategories of Clothing
\& Accessories (b), Women Clothing (c), and Movies (d) are the
major fractions. The fact that these three subcategories belong to
the same cluster implies that they are strongly correlated in an
online transaction. For cluster 3, the subcategory of Sports
Trading Cards (e) is dominant. For cluster 4, (f),(g),(h), and (i)
subcategories exhibit a strongly correlation. For cluster 5,
subcategories (j),(k),(l), and (m) are correlated, which represent
the Pop Culture of Collectibles, Computers, Consumer Electronics,
and Movies, respectively. } \label{FIG:CLUSTER_PIC1}
\end{figure*}

To illustrate closeness, we select a cluster $\alpha$ and classify
the items within the cluster into 192 subcategories. The fraction
of items in each subcategory $\mu$ is the closeness $C_{\alpha
\mu}$. For example, Fig.~\ref{FIG:CLUSTER_PIC1} shows the
closenesses for the first five largest clusters. Each strip
represents a cluster obtained from the HA algorithm. For each
strip, the $x$-axis represents 192 subcategories, and the $y$-axis
does the closeness. The bar indicates the closeness. For cluster
1, subcategory (a) exhibits the largest closeness. For cluster 2.
subcategory (c) has the largest closeness, and so on. The
abbreviations for the 18 main categories are as follows: {\em
Antique} stands for antiques and art; {\em Biz}, business and
office; {\em Clothes}, clothing and accessories; {\em Collect},
collectibles; {\em Comp}, computers; {\em Elec}, consumer
electronics; {\em Dolls}, dolls and bears; {\em Home}, home and
garden's; {\em Jewelry}, jewelry, gemstones and watches; {\em
Glass}, pottery and glass; and {\em Estt}, real estate.

\subsection{Correlation matrix}

To quantify the correlation, we adopt the method used for the
clustering analysis for expression data from DNA microarrays. In
this approach, we regard the closeness as the expression level,
subcategories as genes, and clusters as different DNA microarray
experiments~\cite{eisen}.

The correlation matrix element $\rho_{\alpha\beta}$ is defined as
\begin{equation}
\rho_{\alpha\beta}=\frac{\langle
C_{\alpha\mu}C_{\beta\mu}\rangle-\langle C_{\alpha\mu}\rangle
\langle C_{\beta\mu} \rangle }{\sqrt{ \big(\langle
C_{\alpha\mu}^2\rangle-\langle C_{\alpha\mu}
\rangle^2\big)\big(\langle C_{\beta\mu}^2\rangle-\langle
C_{\beta\mu} \rangle^2\big)}},
\end{equation}
where $C_{\alpha\mu}$ represents the closeness of subcategory
$\mu$ ($\mu=1,\dots n=192$) to cluster $\alpha$ ($\alpha=1,\dots
870$) and $\langle \cdots \rangle$ denotes the average over
different clusters indexed by $\mu$.

Based on the correlation matrix, a dendrogram that assembles all
$n=194$ subcategories into a single tree can be constructed. For
this purpose, we use the hierarchical clustering algorithm
introduced by Eisen {\it et al.} \cite{eisen}. We start the tree
construction process by calculating the correlation coefficients
$\{\rho_{\alpha\beta}\}$ with size $192 \times 870$. Next, the
matrix is scanned to identify a pair of subcategories with the
highest value of the correlation coefficient, and the two
subcategories are combined. Thus, a pseudo-subcategory is created,
and its closeness profile is calculated only by averaging
closenesses of the combined subcategories. This is referred to as
the average-linkage clustering method. Then, $n-2$ isolated
subcategories and a pseudo-subcategory remain. The correlation
matrix is updated with these $n-1$ units and the highest
correlation coefficient is found. The process is repeated $n-1$
times until only a single element remains. After these steps, a
dendrogram is constructed in which the height
represents the magnitude of the correlation coefficient.\\

\begin{figure*}[h]
\center \epsfxsize 15cm \epsfbox{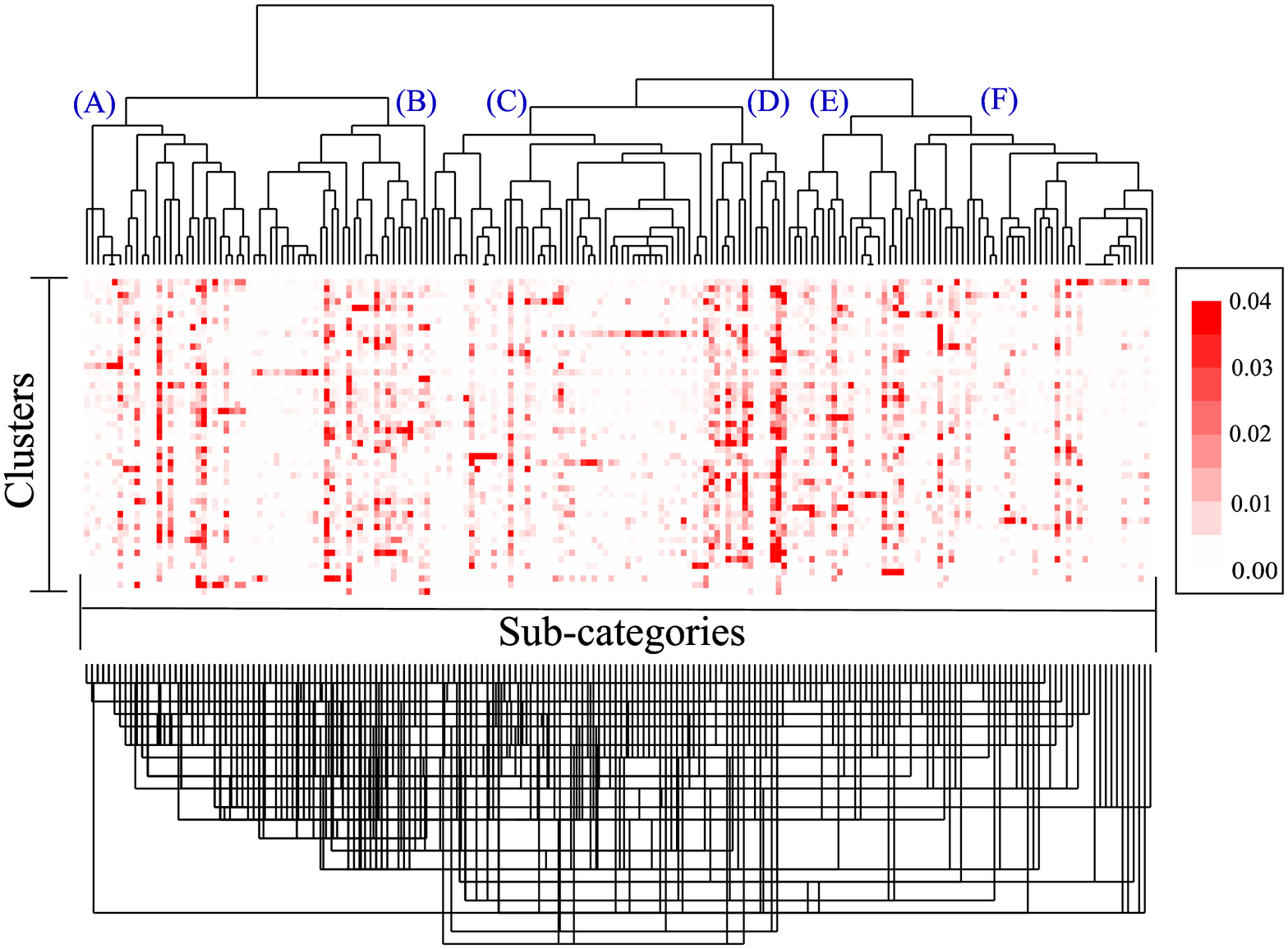} \caption{Upper part:
the dendrogram constructed by using the hierarchical clustering
algorithm for the item network of an eBay online auction.
Subcategories in branches (A)--(F) are explained in the text.
Middle part: the closenesses of each subcategory for different
clusters are shown with various concentrations. Lower part: the
traditional classification scheme of subcategories in the version
where original data were collected. The classification scheme is a
bilayer structure comprising 18 categories and 192 subcategories.
For visual clarity, however, the bilayer structure is shown in a
multilayer manner. For comparison, the subcategories are in the
same order as that used in the upper part. We can easily observe
that the traditional classification scheme is entangled from the
bidder-oriented perspective.} \label{FIG:tax1}
\end{figure*}

\subsection{Rearrangement of subcategories in the dendrogram}

The resulting dendrogram is shown in the upper part of
Fig.~\ref{FIG:tax1}, which is considerably different from the
traditional classification scheme shown in the lower part of this
figure. We discuss the details of the correlations of the
subcategories in the dendrogram. For discussion, we divide the
entire tree structure into six branches, denoted by (A)--(F).

To be specific, branch (A) covers a broad range of different
collectibles. The relationship between the subcategories may be
attributed to collecting manias. Branch (B) mainly covers three
types of subcategories: {\em clothing} and {\em accessories}, {\em
business, office} and {\em industries}, and {\em sports}
categories. Branch (C) consists of three parts: the first part has
{\em antiquary property} and the items used for {\em decorating
homes} and the second part covers very broad kinds of items. The
third is interesting and covers a set of {\em electronic products}
such as computers, cameras, audio players, etc. It also includes
{\em video games} as well as {\em food} and {\em beverages}. At a
glance, one may wonder how these two items are correlated;
however, by considering the fact that some video games maniacs
requires foods and beverages while playing, one can find the
reason. Thus, the dendrogram indeed reflects the bidding patterns
of individual bidders. Branch (D) covers items related to {\em
artistic collections} and {\em hobbies}. Branch (E) covers {\em
books, dolls} for children, etc. Finally, branch (F) mainly covers
collectibles in a wide range from {\em jewelry} to {\em stamps}.

\section{Conclusions and Discussion}
Based on the empirical data collected from the eBay web site, we
have constructed a bipartite network comprising bidders and items.
The bipartite network is converted into two single species of
networks, the bidder and item networks. We measured various
topological properties of each network. Both networks are scale
free in the degree distribution. It is noteworthy that both the
networks are assortatively mixed with regard to the degree
correlation. This fact implies that the active bidders tend to
simultaneously bid for common items; therefore, they are
connected. Accordingly, attractive items are connected via such
active bidders. Next, by applying the hierarchical agglomeration
algorithm, we identified clusters in the bidder and item networks.
The clusters are well separated from each other. Then, we
calculate the correlation matrix between subcategories by using
the information on the fraction of items in each subcategory in a
given cluster. By using this correlation matrix, we construct the
dendrogram, which is different from the traditional classification
scheme. Based on a detailed investigation about the items closely
located in the dendrogram, we find that the dendrogram indeed is
bidder-oriented in an online auction. Therefore, the dendrogram
could be useful for marketing renovation, resulting in an increase in
profits. \\

This work was supported by KRF Grant No.~R14-2002-059-010000-0 of
the ABRL program funded by the Korean government (MOEHRD). \\

\end{document}